\documentclass[reqno]{amsart}
\usepackage{amscd,amsfonts,amssymb}
\usepackage{graphicx}
\usepackage{color}
\usepackage[russian]{babel}
\numberwithin{equation}{section}
\newtheorem{Theorem}{Theorem}
\newtheorem{Lemma}{Lemma}

\theoremstyle{definition}

\newtheorem{proposition}{Proposition}
\theoremstyle{remark}%

\begin{document}
\begin{center}
{\bf Behavior of a boundary of moving volume inside of a smooth
flow of compressible liquid} \vskip2cm

{\it Olga Rozanova} \vskip1cm

Department of Differential Equations,
Mathematics and Mechanics Faculty     \\
Moscow State University                 \\
Glavnoe zdanie GSP-2 Vorobiovy Gory 119992 Moscow
Russia \\
E-mail: {\it rozanova@mech.math.msu.su} \vskip1cm {\bf Abstract.}
\end{center}
{\it The question on expansion of moving volume inside of a smooth
flow of the compressible liquid is under consideration. We find a
condition on initial data such that if it holds, then within a
finite time either the boundary of the moving volume attains a
given neighborhood of a certain point (that do not belong to the
volume initially), or some a priori estimate for the pressure on
the boundary of volume fails.}

\vskip2cm


Let us consider the following system of conservation laws
$$ \rho (\partial_t
{\bf V}+({\bf V},{\bf \nabla})\,{\bf V})= -{\bf\nabla}
P,\eqno(1)$$
$$ \partial_t \rho +
{\rm div}\, (\rho{\bf V})=0\eqno(2)$$
$$ \partial_t S +({\bf
V},{\bf\nabla}) S=0.\eqno(3)$$ The density, velocity vector and
entropy ($\rho, {\bf V}=(V_1,...,V_n)$ and $S,$ respectively) are
unknown. These functions depend on time $t$ and on point
$x=(x_1,...,x_n)\in {\mathbb R}^n.$ Here $P(t,x)$ is the pressure,
$\gamma$ is the adiabatic exponent ($\gamma=const>1$). We consider
(1 -- 3) together with the state equation
$$P=\rho^\gamma e^S.\eqno(4)$$

It follows from  (3) and (4), that for smooth solutions
$$
\partial_t P +({\bf V},{\bf\nabla} P)+\gamma P\, div\,{\bf
V}=0\eqno(5) $$ holds.

We set the Cauchy problem for (1--3), namely,
$$\rho(0,x)=\rho_0(x), \, {\bf V}(0,x)={\bf V}_0(x),\,S(0,x)=S_0(x).
\eqno(6)$$ According to (4), the pressure $P(t,x)$ can be
expressed initially through initial data (6) as
$P(0,x):=P_0(x)=\rho_0^\gamma (x)e^{S_0(x)}.$

 Problem (1--3, 6) has  a solution so smooth as initial data (locally in time),
 that is, for example, if the initial data are of class
$C^1({\mathbb R}^n),$ then there exists such $T_*>0,$ that for
$t\in [0,T],\,T<T_*,$ the solution to system (1--3) will be
classical (\cite{VolpertKhudiaev},\cite{Majda}). The pressure has
the smoothness  $C^1$ as well.

We consider a finite moving volume ${\mathcal V}(t)$(may be,
disconnected), with the smooth boundary $\partial{\mathcal V}(t),$
consisting from the same particles.

We suppose that initially a certain point $x_0$ do not belong to
${\mathcal V}(t).$ Let us set the following question: what
conditions we have to impose on initial data provided they are
known only inside ${\mathcal V}(0),$ to guarantee that within a
time where the flow keeps smoothness, the boundary of given moving
volume will attain a given $\varepsilon$ - neighborhood of point
$x_0$? It is clear that for the answer to  this question it needs
to do some assumptions on the thermodynamic values in the whole
space, not only inside ${\mathcal V}(0).$

It is known that the lose of smoothness signifies that either the
solution itself or its gradient rase without bound
\cite{VolpertKhudiaev},\cite{Majda},\cite {Chemin1}. As the
solution keeps smoothness at $t\in [0,T],\,T<T_*,$ then there
exists $M(T),$ such that
$$\Bigl|\int\limits_{\partial{\mathcal V}(t)}(\frac{\bf x}
{|{\bf x}|},{\bf N}) P(t,x)\,d\Gamma\Bigr|\le
M(T),$$ where $d\Gamma $ is an element of surface
$\partial{\mathcal V}(t).$ The assumption is that we know the constant $M(T)$ in
advance.
\medskip

{\sc Definition} {\,\it We will say that the pressure in the
moment $t$ is {\rm\bf distributed} along the boundary $\partial
{\mathcal V}(t)$ of domain ${\mathcal V}(t)$ {\rm \bf regularly
with the constant $M\ge 0,$ } if}
$$\Bigl|\int\limits_{\partial{\mathcal V}(t)}(\frac{\bf x}
{|{\bf x}|},{\bf N}) P(t,x)\,d\Gamma\Bigr|\le
M.\eqno(7)$$

\medskip

We point out that if the pressure $P(t,x)$ is constant, then the
integral in the left hand side of (7) is equal to zero. When we
set $M$ sufficiently small, we assume that the volume will not
occur in the zone of large gradient of pressure.

Let us recall that equations (1--3, 5) are differential
corollaries of conservation laws (namely, conservation of mass
$m$, impulses and total energy $E$) taking into account the first
law of thermodynamics \cite {Ovs}.
 In our notation
 $$m=\int\limits_{{\mathcal V}(t)}\rho(t,x)\,dx;$$
in the terms of density, velocity and pressure the total energy
has the form
$$E=\int\limits_{{\mathcal V}(t)}(\frac{1}{2}\rho(t,x)|{\bf
V}|^2+\frac{1}{\gamma-1} P(t,x))\,dx.$$

Let us introduce the functional
$$G_\phi(t)=\int\limits_{{\mathcal
V}(t)}\rho(t,x)\phi(|{\bf x}|)\,dx.$$ We denote ${\bf
\sigma}=(\sigma_1,...,\sigma_K)$ the vector with components
$\sigma_k=V_i x_j-V_j x_i,
\,i>j,\,i,j=1,...,n,\,k=1,...,K,\,K={\rm C}_n^2.$

Lemmas 1 and 2 contain some properties of $G_\phi(t),$ that we use
below.

\begin{Lemma} Let us suppose that  $\phi (|{\bf x}|)$
belongs to the class $C^2$ inside ${\mathcal V}(t)$. For classical
solution to system (1), (2), (5) the following equalities take
place:
$$\frac{dG_\phi(t)}{dt}=\int\limits_{{\mathcal V}(t)}
\frac{\phi'(|{\bf x}|)}{|{\bf x}|} ({\bf V},{\bf x})\rho\,dx,$$
$$\frac{d^2 G_\phi(t)}{dt^2}=I_{1,\phi}(t)+I_{2,\phi}(t)+I_{3,\phi}(t)+I_{4,\phi}(t),$$
where
$$I_{1,\phi}(t)=\int\limits_{{\mathcal V}(t)}\frac{\phi''(|{\bf x}|)}{|{\bf
x}|^2} |({\bf V},{\bf x})|^2\rho\,dx,$$
$$I_{2,\phi}(t)=\int\limits_{{\mathcal V}(t)}\frac{\phi'(|{\bf x}|)}{|{\bf
x}|^3} |{\bf \sigma}|^2\rho\,dx,$$
$$I_{3,\phi}(t)=\int\limits_{{\mathcal V}(t)}(\phi''(|{\bf x}|)+
(n-1)\frac{\phi'(|{\bf x}|)}{|{\bf x}|} ) P\,dx,$$
$$I_{4,\phi}(t)=-\int\limits_{\partial{\mathcal V}(t)}
\frac{\phi'(|{\bf x}|)}{|{\bf x}|} ({\bf x},{\bf N}) P\,d\Gamma,$$
where ${\bf N}$ is the outer unit normal to the boundary
$\partial{\mathcal V}(t).$
\end{Lemma}

To prove Lemma 1 we firstly apply the formula of differentiation
with respect to time for the integral taken over a moving volume
\cite{Ovs}, namely,
$$\frac{d}{dt}\int\limits_{{\mathcal V}(t)}f(t,x) dx=
\int\limits_{{\mathcal V}(t)}\bigl(\frac{\partial f(t,x)}{\partial
t}+ {\rm div}(f(t,x){\bf V})\bigr)\,dx,$$  then we use the general
Stokes formula. For example, taking into account (2) we get that
$$\frac{d G_\phi(t)}{dt}=\int\limits_{{\mathcal V}(t)}\rho'_t(t,x)\phi(|{\bf
x}|)\,dx=\int\limits_{{\mathcal V}(t)}\Bigl(-{\rm div}(\rho{\bf
V})\phi(|{\bf x}|)+{\rm div}\bigl(\rho{\bf V}\phi(|{\bf
x}|)\bigl)\Bigr) \,dx=$$$$=\int\limits_{{\mathcal
V}(t)}\bigl(\nabla\phi(|{\bf x}|), {\bf
V}\rho\bigr)\,dx=\int\limits_{{\mathcal V}(t)}\frac{\phi'(|{\bf
x}|)}{|{\bf x}|} ({\bf V},{\bf x})\rho\,dx.$$ $\Box$
\medskip

\begin{Lemma} Let $\phi''(|{\bf x}|)>0.$ Then in the assumptions of Lemma 1
$$\left(\frac{d G_\phi(t)}{dt}\right)^2\le \sup\limits_{{\bf x}\in{\mathcal V}(t)}\frac{(\phi'(|{\bf
x}|))^2} {\phi''(|{\bf x}|)\phi(|{\bf x}|)}\,G_\phi(t)\,
\int\limits_{{\mathcal V}(t)} \frac{\phi''(|{\bf x}|)}{|{\bf
x}|^2}({\bf x,V})^2\rho \,dx\, .$$
\end{Lemma}

To prove this Lemma it is sufficient to apply the H\"older
inequality to the integral $\frac{d G_\phi(t)}{dt}.$ $\Box$

\medskip

Let us choose $\phi(|{\bf x}|)=|{\bf x}|^q,\, q<0.$ We will denote
in this case $G_\phi(t), $ $\frac{d G_\phi(t)}{dt},$
$I_{i,\phi}(t),\,$ $i=1,...,4$ as $G_q(t), F_q(t), I_{i,q}(t),\,$
respectively. We note that for this choice of $\phi(|{\bf x}|)$
the value of $G_q(t)$ is non-negative.

\medskip

To facilitate the formulation of results we introduce the
following denotation:
$$Q_q(t):= \frac{2mE}{1+|q|}
\left(1+\frac{\varepsilon M}{2E}-
|q+n-2|\frac{CG_q^\gamma(t)}{2E}\varepsilon^{-(q\gamma+n(\gamma-1))}
\right), \eqno(8)$$
$$ R_q(t):=\left(|Q_q(t)|\right)^{1/2}
\eqno(9)$$ where
 $C$ is a positive constant, depending on initial data, $n$ and
$\gamma,$ defined below in (14),(15),(16), $\varepsilon>0,\,$
$M\ge 0.$ We will denote ${\rm dist}(\partial{\mathcal V}(t),x_0)$
the distance from the boundary of moving volume to $x_0=0.$

\medskip

\medskip
The main result of the paper is the following theorem:

\begin{Theorem}
Let a finite moving volume ${\mathcal V}(t)$ of compressible
liquid (subject to system (1--3)) with $C^1$ - smooth boundary
$\partial{\mathcal V}(t)$ do not contain a certain point $x_0.$
Suppose that the flow is $C^1$ -- smooth for all
$t\in[0,T],\,T\le\infty,$ and the pressure along the boundary
$\partial{\mathcal V}(t)$ is distributed regularly with a constant
$M$ uniformly in $t$ for $t\in[0,T].$
 Let us suppose also that
$s_0:=\min\limits_{{\mathbb R}^n} S_0(x)>-\infty.$

Choose some real numbers  $q<-n-\frac{2}{\gamma-1} $ and
$\varepsilon,$ where $\varepsilon$ is such that
$\,0<\varepsilon<{\rm dist}(\partial{\mathcal V}(0),x_0).$

Then for all initial data (6) there exists such constant
$\delta\le 0,$ depending on $\varepsilon,
\,q,\,T,\,M,\,n,\,\gamma,m,\,E,\,G_q(0),\,s_0,$ that if initially
$$ \int\limits_{{\mathcal V}(0)}|{\bf x}-x_0|^{q-2}({\bf
V}(0,x),{\bf x}-x_0)\rho_0(x)\,dx<\delta, \eqno(10)
$$
then within a time $t_1$, later then $T,$ the boundary of given
liquid volume will attain  the $\varepsilon$ -- neighborhood of
point $x_0.$

\medskip
More precisely, if
$$ Q_q(0)>0,\, \mbox{then} \quad \delta=
-\varepsilon^{q-1}R_q(0){\rm
cth}\left(\frac{(|q|+1)R_q(0)T}{2\varepsilon m}\right), \eqno(11)
$$
if
$$Q_q(0)=0,\, \mbox{then} \quad\delta=-\frac{\varepsilon^{q}m}{(|q|+1)T},
\eqno(11')
$$
if
$$Q_q(0)<0,\quad T\ge \frac{\pi\varepsilon}{2(|q|+1)R_q(0)}, \,
 \mbox{then} \quad \delta=0,
\eqno(11'')
$$
if
$$
Q_q(0)<0,\quad T<\frac{\pi\varepsilon}{2(|q|+1)R_q(0)},\,
 \mbox{then} \quad\delta=-\varepsilon^{q-1}R_q(0){\rm
ctg}\left(\frac{(|q|+1)R_q(0)T}{\varepsilon m}\right).\eqno(11''')
$$
\end{Theorem}

\medskip

{\sc Remark 1.} Condition (10) requires that the integral in its
left hand side will be negative and, generally speaking,
sufficiently large by modulus. To guarantee this, either the
boundary of volume ${\mathcal V}(0)$ should be sufficiently close
to the point $x_0,$ or the velocity and density should be large
inside the chosen volume. The negativity of integral is connected
with the fact that the vector if velocity inside the volume
initially "in average" forms the obtuse angle with the vector
${\bf x}-x_0.$
\medskip

 Proof of Theorem 1. For the simplicity we take $x_0=0.$
If the theorem statement is not true, then, despite condition
(10), there exists $\varepsilon>0$ such that the neighborhood
$B_\varepsilon^n(0)=\{x\in {\mathbb R}^n\big||{\bf
x}|<\varepsilon\}$ will never belong to the moving volume
${\mathcal V}(t)$, if initially the origin do not belong to
${\mathcal V}(0).$

Since $0\notin {\mathcal V}(t),$ then according to our assumption
$\phi(|{\bf x}|)$ has no singularities inside ${\mathcal V}(t).$

It follows from Lemma 2 that
$$(F_q(t))^2\le \frac{|q|+1}{|q|}
G_\phi(t) \int\limits_{{\mathcal V}(t)} \frac{\phi''(|{\bf
x}|)}{|{\bf x}|^2}({\bf x,V})^2\rho \,dx .\eqno(12)$$

Further we apply Lemma 1 together with  (12) and get that
$$\frac{dF_q(t)}{dt}\ge \frac{|q|+1}{|q|}\frac{F_q(t))^2}{G_q(t)}+
I_{2,q}(t)+I_{3,q}(t)+I_{4,q}(t),\eqno(13)$$ with
$$I_{2,q}(t)=q\int\limits_{{\mathcal V}(t)}|{\bf x}|^{q-4}
|{\bf \sigma}|^2\rho\,dx,$$
$$I_{3,q}(t)=q(q+n-2)\int\limits_{{\mathcal V}(t)}|{\bf x}|^{q-2} P\,dx,$$
$$I_{4,q}(t)=-q\int\limits_{\partial{\mathcal V}(t)}|{\bf x}|^{q-1}(\frac{\bf x}
{|{\bf x}|},{\bf N}) P\,d\Gamma.$$ The first item in this sum is
non-negative, the second item is non-positive, the tird one is
positive provided $q<2-n$ and non-positive otherwise, the fourth
one can have any sight in dependence on the form of the boundary
${\mathcal V}(t).$

Let us estimate these items. We obtain that
$$|I_{2,q}(t)|\le 2 |q| \varepsilon^{q-2}E,$$
$$|I_{4,q}(t)|\le q
\varepsilon^{q-1}M.$$ To estimate
 $I_{3,q}(t)$ we need the following Lemma.

\medskip
\begin{Lemma} Let $q<-n-\frac{2}{\gamma-1}$. Then, if
$B_\varepsilon^n(0)\notin {\mathcal V}(t)$,  the following
estimate holds:
$$\int\limits_{{\mathcal V}(t)}|{\bf x}|^{q-2}
 \rho^\gamma\,dx\ge C_1
 G_q^\gamma(t)\varepsilon^{-((q+n)(\gamma-1)+2)},
$$
here the positive constant $C_1$ depends on $\gamma,q,n.$
\end{Lemma}

Let us prove Lemma 3. According to the H\"older inequality we have
that
$$G_q(t)=
 \int\limits_{{\mathcal V}(t)}|{\bf x}|^q
 \rho\,dx=
 \int\limits_{{\mathcal V}(t)}|{\bf x}|^{q}
 \rho \frac{|{\bf x}|^{\frac{q-2}{\gamma}}}
 {|{\bf x}|^{\frac{q-2}{\gamma}}}\le$$
 $$
 \le \left(\int\limits_{{\mathcal V}(t)}|{\bf x}|^{q-2} \rho^\gamma\,
 dx\right)^{\frac{1}{\gamma}}
 \left(\int\limits_{{\mathcal V}(t)}
 |{\bf x}|^{\frac{2+q(\gamma-1)}{\gamma-1}}
 \,dx\right)
 ^{\frac{\gamma-1}{\gamma}}\le$$
 $$\le C_2 \left(\int\limits_{{\mathcal V}(t)}|{\bf x}|^{q-2} \rho^\gamma\,
 dx\right)^{\frac{1}{\gamma}},$$
 with $$C_2=\left(\int\limits_{{\mathbb R}^n\backslash B_\varepsilon^n (0)}
 |{\bf x}|^{\frac{2+q(\gamma-1)}{\gamma-1}}
 \,dx\right)
 ^{\frac{\gamma-1}{\gamma}}=$$
 $$=\left(\frac{\sigma_n(1-\gamma)}{(q+n)(\gamma-1)+2}\right)^
 {\frac{\gamma-1}{\gamma}}\varepsilon^{\frac{(q+n)(\gamma-1)+2}
 {\gamma}}.$$
 The integral converges provided  $q$ satisfies to restrictions
 from the  statement of Lemma 3.

Thus, we denote $$
 C_1:=\left(\frac{\sigma_n(1-\gamma)}{(q+n)(\gamma-1)+2}\right)^
 {1-\gamma},\eqno(14)$$
 and obtain the result of Lemma 3 after elementary calculations. $\Box$

\medskip

We continue the proof of the Theorem. Now we can estimate the
integral  $I_{3,q}(t)$ from below. First of all, we point out that
it is positive if $q$ satisfies the restrictions from the Theorem
statement, since $q<-n-\frac{2}{\gamma-1}$ implies $q<2-n.$
Further, according to equation (3) the entropy is conserved along
trajectories of particles. Therefore $S(t,x)\ge
\min\limits_{{\mathbb R}^n} S_0(x)=s_0.$ Let us denote
$$C_3=e^{s_0}.\eqno(15)$$
Since we assume that $s_0>-\infty,$ then $C_3>0.$ It follows from
Lemma 3 that for $q$ from the Theorem statement
$$I_{3,q}(t)\ge C|q||q+n-2|
 G_q^\gamma(t)\varepsilon^{-((q+n)(\gamma-1)+2)},
$$
where the constant $C$ is defined as follows:
$$C:=C_1 C_3.\eqno(16)$$

Besides we have $G_q(t)\le \varepsilon ^q m.$

Taking into account all estimates obtained, we get from (13)
$$\frac{d}{dt}F_q(t)\ge \frac{|q|+1}{|q|}\frac{F^2_q(t)}{\varepsilon^q m}-
2E|q|\varepsilon^{q-2}+|q||q+n-2|CG_q^\gamma(t)\varepsilon
^{-((q+n)(\gamma-1)+2)}-|q|M\varepsilon^{q-1}=$$
$$=\frac{|q|+1}{|q|\varepsilon^q m}(F^2_q(t)-q^2
\varepsilon^{2q-2}Q_q(t)),\eqno(17)$$ where $Q_q(t)$ is defined in
(8).

We consider firstly the case  $Q_q(0)>0.$

Assume that initially condition (10) is satisfied with the value
of $\delta,$ given by (11), that is
$$F_q(0)>|q|\varepsilon^{q-1}R_q(0){\rm
cth}(\frac{\lambda T}{2}),\eqno(18)$$ where
$\lambda:=\frac{(1+|q|)R_q(0)}{\varepsilon m},$ the function
$R_q(0)$ is defined in (9).

As follows from (17), under this condition within a time  $t\in
[0,\tau),\,\tau>0$ the function $F_q(t),$ and, consequently,
$G_q(t)$ will increase. The function $Q_q(t)$ will decrease,
inversely, as one can see from its expression. Thus, in the moment
$t=\tau$ the inequality
$$\frac{d F_q(t)}{dt}(\tau)\ge \frac{|q|+1}{|q|\varepsilon^q m}(F^2_q(t)-q^2
\varepsilon^{2q-2}Q_q(0))$$ holds. Acting analogously, we can
prolong the interval of decreasing of $Q_q(t)$ till $t\in
[0,2\tau).$ By means of this procedure, we can in finite number of
steps attain the time $T.$ So, for all  $t\in [0,T)$ the
inequality
$$\frac{d F_q(t)}{dt}\ge \frac{|q|+1}{|q|\varepsilon^q m}(F^2_q(t)-q^2
\varepsilon^{2q-2}Q_q(0))\eqno(19)$$ takes place.

Let us note that condition (18) signifies that inequalities
$$F_q(t)>|q|\varepsilon^{q-1}R_q(0){\rm cth} (\lambda T)>
|q|\varepsilon^{q-1}R_q(0)\eqno(20)$$ hold.

 Integrating (19), we obtain that
 $$F_q(t)\ge \frac{|q|\varepsilon^{q-1}R_q(1+{\mathcal
K}e^{\lambda t})}{1-{\mathcal K}e^{\lambda t}},$$ where $
{\mathcal K}=\displaystyle\frac{F_q(0)-|q|\varepsilon^{q-1}R_q(0)}
{F_q(0)+|q|\varepsilon^{q-1}R_q(0)}=const.$ If (20) holds, then
${\mathcal K}<1$ and $F_q(t)$ becomes unbounded within a finite
time $t_*\le T_1=\frac{1}{\lambda}\ln \frac{1}{\mathcal K}.$

However, as follows from the H\"older inequality,  $|F_q(t)|$ is
bounded, namely,
$$|F_q(t)|\le |q| \varepsilon^{q-1}\sqrt{2mE}.\eqno(21)$$
Thus, if we show that $T_1<T, $ then we will go to a contradiction
with the assumption on smoothness of solution up to the moment
$T.$

Indeed, we solve inequality
$$\frac{1}{\lambda}\ln
\frac{1}{\mathcal K}<T,$$ and obtain that it holds if the
condition
$$F_q(0)>|q|\varepsilon
^{q-1}R_q (0)\frac{e^{\lambda T}+1}{e^{\lambda T}-1}$$ holds, it
is the same as (18) (or (10)).

Now let $Q_q(0)=0.$ Analogously to the preceding case on can show
that if conditions (10) and (11') hold, then the functions
$F_q(t)$ and $G_q(t)$ increase, therefore $Q_q(t)$ decreases. So,
we have from (17) that
$$\frac{dF_q(t)}{dt}\ge \frac{|q|+1}{|q|\varepsilon^q m}F^2_q(t).\eqno(22)$$
Integrating (22) we obtain that for all $F_q(0)>0$ within a time
later then $T_2=\frac{|q|\varepsilon^q m}{|q-1|F_q(0)},$ the
function $F_q(t)$ become unbounded. Condition (10) is the
requirement imposed on $F_q(0)$ for the implementation of
inequality $T_2<T.$

At last, if  $Q_q(0)<0,$ then for any initial data such that
$F_q(0)\ge 0,$ functions $F_q(t)$ and $G_q(t)$ increase, $Q_q(t)$
decreases, therefore  from (17) we get
$$\frac{dF_q(t)}{dt}\ge \frac{|q|+1}{|q|\varepsilon^q m}(F^2_q(t)+q^2
\varepsilon^{2q-2}|Q_q(0)|).\eqno(23)$$

Integrating (23) we obtain $${\rm
arctg}\frac{F_q(t)}{|q|\varepsilon^{q-1}R_q(0)}\ge {\rm
arctg}\frac{F_q(0)}{|q|\varepsilon^{q-1}R_q(0)}+
\frac{|q|+1|R_q(0)}{\varepsilon m}t, $$ and one can conclude that
function $F_q(t)$ becomes unbounded within a time later then
$$T_3=\left(\frac{\pi}{2}-{\rm
arctg}\frac{F_q(0)}{|q|\varepsilon^{q-1}R_q(0)}\right)
\frac{\varepsilon m}{(|q|+1)R_q(0)}.$$ If $T\ge
\frac{\pi\varepsilon}{2(|q|+1)R_q(0)}$ (this is condition (11'')),
then $T>T_3.$  If $T<\frac{\pi\varepsilon}{2(|q|+1)R_q(0)},$ then
for the implementation of inequality $T>T_*,$ the value of
$F_q(0)$ should be bounded from below by the constant $|q|\delta,$
with $\delta$ indicated in (11'''), that is (10) should hold.

So, Theorem 1 is completely proved.
 $\Box$

\medskip

\medskip
However, it needs to point out that it may occur  that we cannot
find initial data and constants $M\ge 0,\, \varepsilon>0,\,T>0,\,$
$q,$ such that (10) holds. The following proposition concerns with
a necessary condition of  implementation of (10) for the case
$Q_q(0)>0$.

\medskip

\medskip
\begin{proposition} {\it Let us suppose that the initial data and the constants
$M\ge 0,\, \varepsilon>0,\,$ $q<-n-\frac{2}{\gamma-1}$ are such
that
$$1+\frac{\varepsilon
M}{2E}-\frac{CG_q(0)}{2E}\varepsilon^{-(2+(q+n)(\gamma-1))}>0,
$$
(that is $Q_q(0)>0$). Let also condition (10) hold.

Then these constants and the time $T,$ when the flow is assumed
smooth, obey the following inequality
$${\rm
cth}\left(\frac{(|q|+1)R_q(0)T}{2\varepsilon
m}\right)<\frac{\sqrt{2mE}}{R_q(0)}.\eqno(24)$$}
\end{proposition}
\medskip
To prove the proposition we note that inequality (21) should be
true at the zero moment of time, too. Then, taking into account
(18), we get the following two-sided inequality:
$$|q|  \varepsilon^{q-1}R_q(0){\rm
cth}\left(\frac{(|q|+1)R_q(0)T}{2\varepsilon m}\right)<F_q(t)\le
|q|{\rm dist(\partial{\mathcal V}(0),x_0)}^{q-1}\sqrt{2mE}.$$
Comparing the left and right sides of the last inequality, we
obtain
$$\frac{1}{\sqrt{2mE}}{\rm
cth}\left(\frac{(|q|+1)R_q(0)T}{2\varepsilon m}\right)R_q(0)<
\left(\frac{\varepsilon}{dist(\partial{\mathcal
V}(0),x_0)}\right)^{1-q}<1,$$ it signifies (24). $\Box$

\medskip
{\sc Remark 2.} Necessary conditions, analogous to (24), for the
case $Q_q(0)=0$ and $Q_q<0,
\,T<\frac{\pi\varepsilon}{2(|q|+1)R_q(0)}$ are
$$
\frac{\varepsilon}{(|q|+1)T}\sqrt{\frac{m}{2E}}<1 \eqno(25)
$$ and
$${\rm
ctg}\left(\frac{(|q|+1)R_q(0)T}{\varepsilon
m}\right)<\frac{2mE}{R_q(0)},$$ respectively.

{\sc Remark 3.} Let us analyze inequality (24). We denote
$\phi_1(R_q(0)):= {\rm
cth}\left(\frac{(|q|+1)R_q(0)T}{2\varepsilon m}\right),\,$
$\phi_2(R_q(0)):=\frac{\sqrt{2mE}}{R_q(0)}.$ First of all, (24)
implies an upper bound for $Q_q(0),$  since $\phi_1(R_q(0))\to 1,$
and $\phi_2(R_q(0))\to 0,$ as $R_q(0)\to\infty.$ For small
$R_q(0)$ $(R_q(0)\to 0)$ inequality (24) signifies that
$$1+\frac{2\varepsilon m}{(1+|q|)T
R_q(0)}<\frac{\sqrt{2mE}}{R_q(0)},$$ or
$$R_q(0)<\sqrt{2mE}-\frac{2\varepsilon m}{(1+|q|)T}.$$
The last inequality can be true only if
$$\frac{\varepsilon}{T}<\frac{(1+|q|)\sqrt{E}}{\sqrt{2m}},$$
it implies inequality (25), that is natural.

\medskip

{\sc Example.} Let us consider the simplest situation, where the
velocity of flow is a constant vector, the density and pressure
are constants. We suppose that the point $x_0$ is situated with
respect to the volume ${\mathcal V}(0)$ such that $({\bf
V}(0,x),{\bf x}-x_0)<0.$ It is easy to see that in this case the
left hand side of (10) is negative. Evidently that in our
situation $T=\infty.$
 Above we  pointed out  that at a constant pressure we can set $M=0.$

We restrict ourself by situations where $Q_q(0)<0.$ In this case,
as follows from (11''), if the left hand side of  (10) is
negative, then the moving volume will attain the $\varepsilon$ --
neighborhood of point $x_0.$

The calculation shows that for constant values of velocity,
density and pressure
$$
\left(1+\frac{\varepsilon M}{2E}-
|q+n-2|\frac{CG_q^\gamma(t)}{2E}\varepsilon^{-(q\gamma+n(\gamma-1))}
\right)\le 1-A |q|^\gamma
\left(\frac{\varepsilon}{d_1}\right)^{\gamma |q|}
\varepsilon^{-n(\gamma-1)},$$ where a positive constant
 $A$ depends only on $\rho_0,\,P_0,\,{\bf
V}_0,\,n,\,\gamma,$ and $d_1:=\displaystyle \sup\limits_{x\in
{\mathcal V}(0)}|x-x_0|.$ It is clear that
$\frac{\varepsilon}{d_1}\le \frac{\varepsilon}{{\rm
dist}(\partial{\mathcal V}(0))}< 1.$ Therefore, if we require
condition
$$
A |q|^\gamma \left(\frac{\varepsilon}{d_1}\right)^{\gamma |q|}
\varepsilon^{-n(\gamma-1)}>1,\eqno(26)$$ then $Q_q(0)$ will be
negative. Condition (26) signifies
$$\ln\varepsilon>\frac{1}{\gamma |q|-n(\gamma-1)}
\ln\left(\frac{d_1^{\gamma |q|}}{A|q|^\gamma}\right),$$ that is
$\varepsilon$ is sufficiently large.

From the other side, condition (26) signifies that if the ratio
$\frac{\varepsilon}{d_1}$ is not very small, that is the moving
volume is close to the $\varepsilon$ -- neighborhood of point
$x_0,$ then the value in the left hand side of inequality (26) can
be large for small $\varepsilon,$ too.

So, if we do not want to miss in the progressive movement the
 $\varepsilon$
-- neighborhood of the point  $x_0,$ then we need to require
initially the volume under consideration to be sufficiently close
to this neighborhood or the neighborhood to be sufficiently large.
We should get the same conclusion from elementary geometric
considerations.

\medskip

{\sc Remark 4.} Functionals $G_q(t)$  can be used for another
purposes. In particular, at $q=2$ it is possible to apply them for
investigation of properties of solutions to equations of
compressible fluid with forcing in several dimension, namely, for
the search of sufficient conditions of a loss of smoothness by
classical solutions \cite{Sideris},\cite{RozDU},\cite{RozSing}.
Besides, by means of these functionals one can find asymptotics of
the kinetic and potential components of total energy for a smooth
flow of the compressible fluid \cite{Chemin}, \cite{RozFAO}, and
construct classes of exact solutions for such system of equations,
as well \cite{RozPas},\cite{RozJMS},\cite{RozNova}.


\medskip\medskip

\begin{thebibliography}{99}

\bibitem{VolpertKhudiaev} A.I.Volpert, S.I.Khudiaev, \it On the
Cauchy problem for composite systems of nonlinear equations. \rm
Mat.Sbornik {\bf 87}(1972), N4, 504--528.


\bibitem{Majda} A.Majda,{\it Compressible fluid flow and systems
of conservation laws in several space variables,} \rm
Appl.Math.Sci.  {\bf 53}(1984), 1--159.

\bibitem{Chemin1}
J.-Y.Chemin,{\it Remarque sur l'apparition de singularit$\acute
e$s dans des ecoulements euleriens compressibles}, Comm.
Math.Phys. {\bf 133}(1990), 323-329.

\bibitem{Ovs}L.V.Ovsyannikov, {\it Lectures on the fundamentals of gas dynamics.}.-
Moskva: "Nauka". 368 p. (1981).


\bibitem{Sideris} T.C.Sideris, \it
Formation of singularities in three-dimensional compressible
fluids, \rm Comm.Math.Phys. {\bf 101} (1985), 475--485.

\bibitem{RozDU} O.S.Rozanova,
{\it Generation of singularities of compactly supported solutions
of the Euler equations on a rotated plane.} Differ. Equat.,
\textbf {34}(8) (1998),  1118-1123.

\bibitem{RozSing}О.С.Розанова, {\it Development of singularities
for the compressible fluid equations with external force in
several dimensions}(to appear in J.Math.Sci.), e-Print
archive:http://arxiv.org/math.AP/0411652 math.AP/0411652




\bibitem{Chemin}J.-Y.Chemin,
\it Dynamique  des  gaz  \`a  masse  totale finie, \rm
   Asymptotic Analysis. {\bf 3}(1990), 215-220.

\bibitem{RozFAO}O.S.Rozanova,\it
Energy balance  in a Model of the Dynamics of a Two-Dimensional
Baroclinic Atmosphere,
  \rm Izvestiya, Atmospheric and Oceanic Physics , {\bf 34}(1998),
   N 6, 738--744.

\bibitem{RozPas}
O.S.Rozanova,{\it On classes of globally smooth solutions to the
Euler equations in several dimensions} in: Hyperbolic problems:
Theory, Numerics, Applications. Proceedings of 9th International
Conference in Pasadena, Caltech, March 25-29,2002,
pp.861-871.Springer,2003.

\bibitem{RozJMS}
O.S.Rozanova, {\it Application of integral functionals to the
study of the properties of solutions to the Euler equations on
riemannian manifolds}, J.Math.Sci. {\bf 117}(5)(2003), 4551--4584.

\bibitem{RozNova}
O.S.Rozanova,{\it Classes of smooth solutions to
multidimensionalbalace law of gas gynamic type on riemannian
manifolds,} in: "Trends in Mathematical Physics Research" ed.C.V.
Benton, New York, Nova Science Publishers, Inc.
  pp. 155-204, 2004.

\end{thebibliography}
\end{document}